# Analysis of water ice in nanoporous copper needles using cryo atom probe tomography

Levi Tegg [a,b]*, Ingrid E. McCarroll [c], Se-Ho Kim [c,d], Renelle Dubosq [c], Eric V. Woods [c], Ayman A. El-Zoka [c,e], Baptiste Gault [c,e], Julie M. Cairney [a,b]*

[a] School of Aerospace, Mechanical and Mechatronic Engineering, The University of Sydney, Sydney, NSW 2006, Australia

[b] Australian Centre for Microscopy and Microanalysis, The University of Sydney, Sydney, NSW 2006, Australia

[c] Max-Planck-Institut für Eisenforschung, Max-Planck-Straße 1, 40237 Düsseldorf, Germany

[d] *now at* Department of Materials Science and Engineering, Korea University, Seoul 02841, Republic of Korea

[e] Department of Materials, Royal School of Mines, Imperial College London, SW7 2AZ London, United Kingdom

*Corresponding authors: Levi.Tegg@sydney.edu.au, Julie.Cairney@sydney.edu.au.

## Abstract

The application of atom probe tomography (APT) to frozen liquids is limited by difficulties in specimen preparation. Here, we report on the use of nanoporous Cu needles as a physical framework to hold water ice for investigation using APT. Nanoporous Cu needles are prepared by the electropolishing and dealloying of Cu-Mn matchstick precursors. Cryogenic scanning electron microscopy and focused-ion beam milling reveal a hierarchical, dendritic, highly-wettable microstructure. The atom probe mass spectrum is dominated by peaks of $Cu^+$ and $H(H_2O)_n^+$ up to $n \leq 3$, and the reconstructed volume shows the protrusion of a Cu ligament into an ice-filled pore. The continuous Cu ligament network electrically connects the apex to the cryostage, leading to enhanced electric field at the apex and increased cooling, both of which simplify the mass spectrum compared to previous reports.

## 1   Introduction

Atom probe tomography (APT) is an analytical technique that provides three-dimensional chemical information at sub-nanometre resolution [1–3]. Stringent requirements on specimen geometry and conductivity have historically restricted application of APT to metals and alloys: specimens should be stable under ultra-high vacuum, be electrically conductive, relatively homogenous and non-porous, and conical with an apex diameter of < 200 nm [2]. The development of the laser-pulsed atom probe [4,5] and focused-ion beam (FIB) sample preparation techniques [6,7] have enabled the analysis of less-conductive samples, such as semiconductors [8–10] and geological materials [11,12]. However, the specimen geometry and conductivity requirements still complicate the analysis of soft matter, such as frozen liquids, polymers, and biological materials. The analytical information that APT provides could complement data from powerful new techniques such as cryogenic transmission electron microscopy [13,14] that allow for the structural determination of proteins or viruses [15,16]. Specifically, the atom probe allows 3D compositional information to be obtained, and has the potential to reveal how these structures are positioned when they are embedded in cellular systems. For these reasons, there is great value in developing specimen preparation workflows for the APT analysis of soft matter.

Several groups have recently reported progress in this area [17–22]. Early studies involved dipping pre-sharpened W tips into an analyte solution before plunge-freezing [23,24]. More recent developments include resin-mounting with FIB milling and lift-out [25], encapsulating aqueous solutions onto a W tip using graphene [16,26–29], using FIB milling and lift-out to observe materials contained in glass [15,30,31], to "stand-out" lamellae by ion-induced bending [32], and milling liquids which have been cast and frozen onto substrates of nanoporous Au [33,34] or microstructured W surfaces [35–38]. These techniques have enabled the study of frozen water [26,33–35], solutions and suspensions of alkali salts [24,33], ferritin [16,25], and of other liquids [36–38]. Each of these studies present a novel approach to analysing soft matter in APT, and provide important insights into the field evaporation properties of water and soft matter [29,33,35,39].

Nanoporous metals have some advantages over other liquid-supporting substrates [21], such as straightforward fabrication [40] and high wettability [41]. Nanoporous metals consist of a network of nanoscale ligaments and pores, which may be randomly distributed and interconnected or patterned and highly-ordered [40]. Several preparation techniques have been reported [42], but one common method is the acid-etching, or "dealloying", of a homogenous binary alloy where one component is significantly more noble than the other [40]. The prototype system is nanoporous Au from the dealloying of an AuAg solid-solution by nitric acid [43], but advances in synthesis have extended the range of metals which can be made nanoporous by dealloying [44–46]. This provides an opportunity to develop substrates which overcome the shortcomings of nanoporous Au (high evaporation field [47] and high cost), and allow for control over the hydrophilicity and the strength of the solid-liquid interface.

Several factors need to be considered in designing a nanoporous metal substrate. The material should be low cost, processable into needle-shaped specimens [2], have potentials suitable for dealloying [40], be chemically inert, have few isotopes, and have an evaporation field similar to that of water ($\approx$ 10 V/nm [39]) to minimise local magnification [33]. In this work, we report on the use of nanoporous Cu as a support for studying water ice in the atom probe. We have selected Cu as the nanoporous metal despite its modest evaporation field (30 V/nm [47]) as it satisfies all of the other criteria listed above [48]. In contrast with previous reports which use flat nanoporous substrates [33,34], we have use needle-shaped specimens.

# 2 Methods

## 2.1 Alloy and needle preparation

Cu and Mn chips (Sigma Aldrich, 99%) were mixed in a 30:70 atomic ratio, heated to ≈1300 °C using an induction furnace, and cast into a graphite mould. A section of the as-cast $Cu_{0.3}Mn_{0.7}$ alloy was polished using SiC paper and chemo-mechanical oxide polishing solution. Scanning electron microscopy (SEM) and energy-dispersive X-ray spectroscopy (EDS) were performed on the polished section using a Zeiss Sigma HD with a 20 kV incident electron beam.

A general overview of the atom probe specimen preparation workflow is shown in Figure 1. Matchstick-shaped sections were cut from the as-cast billet using a diamond saw. The matchsticks were electropolished into needles of ≈10 μm diameter using a solution of 75 vol.% $H_3PO_4$ in $H_2O$ with a DC voltage of 3–7 V against a copper electrode. Needles were rinsed with distilled $H_2O$ and methanol, crimped with a Cu sleeve, then stored in air.

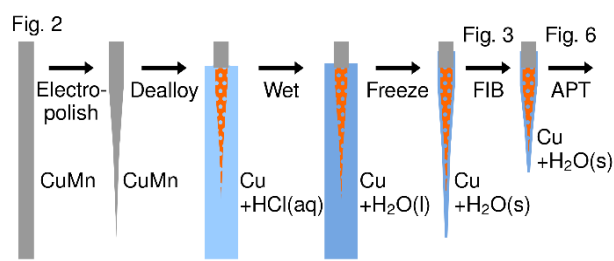

Figure 1: A overview of the preparation of ice-loaded nanoporous metal tips starting from a matchstick-shaped section of the parent alloy. The top row of labels indicate figures showing micrograph of the specimen at selected steps in the specimen preparation procedure. Nominal specimen composition is indicated by the labels in the lower half of the figure, and "CuMn" here represents $Cu_{0.3}Mn_{0.7}$.

The needles were mounted into Cameca low-profile VCTM-compatible specimen pucks ("cryo-pucks"). To produce the nanoporous structure, the ends of the needles were immersed in 0.2 M HCl solution in distilled $H_2O$ at room temperature for 8 min [48,49]. The needles were then soaked in distilled $H_2O$ for >15 min inside a $N_2$ glovebox [18,19,50] to dilute and exchange the dealloying acid. Care was taken to ensure the tip was never dried and the puck was never wet. The wet tip was removed from the $H_2O$, swiftly plunge-frozen in liquid nitrogen, and then transferred to a Ferrovac ultra-high vacuum cryotransfer suitcase.

The atom probe data shown in this manuscript was collected on a specimen which underwent additional processing steps to what is described above. It was electropolished, dealloyed in 0.2 M HCl for 5 min, dried in air, then FIB-milled at room temperature to remove the apex and produce a truncated needle, then dealloyed in 0.2 M HCl again, rinsed in $H_2O$, and plunge-frozen before cryo-transfer. These additional steps may not be necessary for successful sample preparation, and similar (though smaller) atom probe datasets have been collected where these extra steps were not taken.

## 2.2 Focused ion-beam milling

Following plunge-freezing in the $N_2$ glovebox, samples were transferred from the Ferrovac suitcase to a FEI Helios Xe plasma dual-beam focused-ion beam (PFIB) under cryogenic and high-vacuum conditions. The specimens were annular-milled at ≈ 80 K to produce a sharp tip suitable for APT. Milling was performed initially using a voltage of 30 kV and a current of 15 nA, and was gradually reduced to 0.1 nA beam current. High-resolution electron-beam imaging in the later stages of milling was avoided to minimise sublimation of ice from the specimen apex. After milling, the sample was transferred back to the Ferrovac suitcase under cryogenic and high-vacuum conditions.

## 2.3 Atom probe tomography

Specimens were transferred from the Ferrovac suitcase into a LEAP 5000 XS [18] under cryogenic and ultra-high vacuum conditions. The atom probe data shown here were collected at temperature 50 K, pressure $< 8 \times 10^{-12}$ Torr, with a pulse frequency of 125 kHz, a laser pulse energy of 50 pJ, a laser wavelength of 355 nm, and a target detection rate of 1% (0.01 ion/pulse). Similar experiments have been performed under other acquisition conditions by using a LEAP 4000 X Si [50], though their results are not shown here. Data reconstruction was performed using the reconstruction wizard in AP Suite 6.3. The voltage and flight-path corrections provided within the IVAS reconstruction wizard were used to calibrate the time-of-flight spectrum. Without accurate electron micrographs or crystallographic poles on the detector, reconstruction parameters were chosen arbitrarily to reproduce a geometry similar to that expected of a nanoporous metal. This means the distances reported in this manuscript should be considered approximate. The fixed-shank angle was 10°, the image compression factor was 1.10, the field factor was 3.3, and the detector efficiency 80%. Analysis of the reconstructed volume was performed using the IVAS module of AP Suite 6.3.

## 3 Results

### 3.1 Alloy microstructure

The microstructure of the parent alloy significantly impacts the nanostructure of the subsequent nanoporous metal [40,44,45,49], so the microstructure of the as-cast $Cu_{0.3}Mn_{0.7}$ was studied using SEM and EDS. Figure 2(a) shows an Everhart-Thornley secondary electron (ETSE) image of a polished surface. EDS mapping shows the distribution of (b) Mn, (c) Cu and (d) O across the surface. This alloy consists of Mn-rich γ–CuMn [51] dendrites in a Mn-poor γ-CuMn matrix, as well as some MnO inclusions. This microstructure results from coring as the alloy slowly cools after casting [49,52–54]. The oxide polishing solution used here has preferentially-oxidised the Mn-rich dendrites, which are more chemically reactive than the Cu-rich matrix. This difference in reactivity leads to challenges in chemo-mechanical polishing and electropolishing. Similarly, the dealloying of the Mn-rich dendrites is more rapid than of the surrounding matrix. Although this complicates the sample preparation, the hierarchical porosity appears to improve wetting and retention of $H_2O$ in the structure [49,55]. Some specimens were prepared from alloy sections homogenised at 900°C for 24 h and quenched [54] to remove the dendritic microstructure, but less ice was found inside these specimens following cryotransfer.

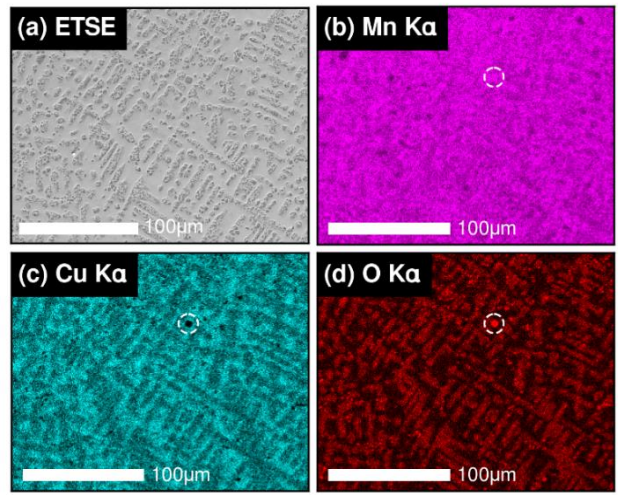

Figure 2: SEM and EDS analysis of the as-cast $Cu_{0.3}Mn_{0.7}$ alloy. (a) Everhart-Thornley secondary electron (ETSE) image of the polished surface, and (b) Mn Kα, (c) Cu Kα and (d) O Kα X-ray intensity maps of the same field of view. The circle annotation shows a manganese oxide inclusion.

## 3.2 Needle microstructure

The microstructure of the bulk alloy is also present in the electropolished and dealloyed needles, and this affects the strategy used to FIB-mill the final tip. Figure 3(a) shows an ETSE image of a specimen which is mid-way through the cryogenic annular milling procedure. Three regions are observed when as-cast samples are dealloyed: nanoporous Cu regions, ice-filled cracks, and pristine bulk γ-CuMn. Redeposition from the ion beam milling appears at the end of the tip, but this is removed later in the annular milling procedure. X-ray spectra from EDS of similar tips are shown in Figure 3(b). From these spectra, the Cu:Mn atomic ratio is ≈95:5 in the nanoporous Cu region and ≈50:50 in the γ-CuMn region, similar to what is expected from coring and dealloying of a 30:70 melt [49,51]. The ice-filled cracks result from the rapid dealloying and disintegration of the Mn-rich dendrites. The γ-CuMn regions have lower Mn content and therefore dealloy more slowly, and have not completely dealloyed before this sample was rinsed and plunge-frozen.

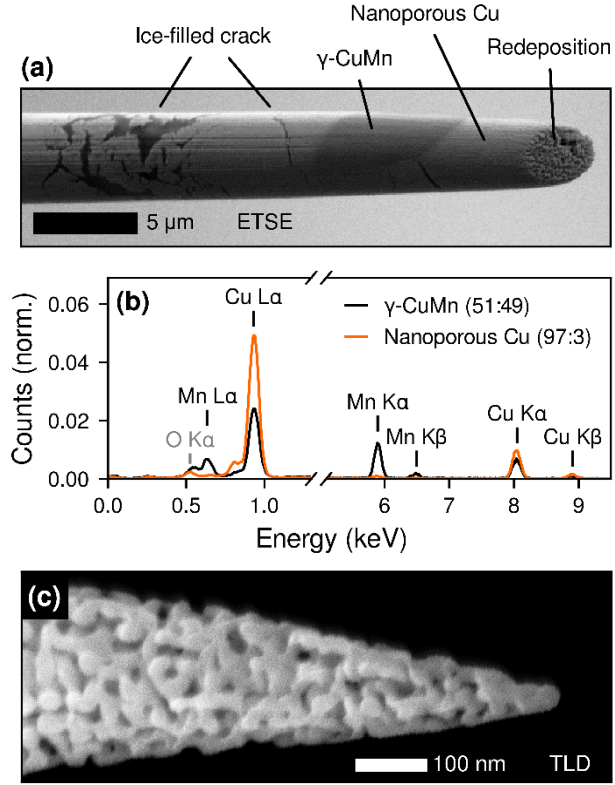

Figure 3: (a) ETSE image of a tip of nanoporous Cu loaded with H$_2$O, imaged under cryogenic conditions (≈ 80 K) and mid-way through the plasma focused-ion beam (PFIB) annular milling procedure. The three main regions, plus redeposition, are labelled. This sample was prepared by dealloying an electropolished as-cast Cu$_{0.3}$Mn$_{0.7}$ sample in 0.2 M HCl for 5 min. (b) EDS X-ray spectra collected from a γ-CuMn region and a nanoporous Cu region in a different sample to those shown in (a) or (c). The ratios in the legend indicate the measured Cu:Mn ratio from each spectrum. Note the split horizontal axis. (c) Through-lens detector (TLD) secondary electron image dry nanoporous Cu tip (i.e. without any H$_2$O) following final-stage PFIB annular milling, illustrating the ideal specimen geometry.

Figure 3(c) shows a tip of nanoporous Cu without ice to illustrate the ideal geometry of the final specimen: Cu ligaments of diameter ≈20–40 nm with pores of diameter ≈15–30 nm [48,49]. The electron beam of the SEM sublimates ice even under cryogenic conditions [21], thus confirming the presence of ice in the nanoporous Cu region by SEM has not been performed. Instead, we have assumed that the presence of ice in the large cracks indicates that ice is also present in the nearby nanopores. The specimen preparation strategy is thus to ensure large ice-filled cracks are intact and somewhat near the specimen apex. The proximal nanopores, which are assumed to also contain ice, are ultimately positioned within the top few hundred nanometres of the specimen by careful FIB annular milling.

### 3.3 Atom probe tomography

Following the specimen preparation strategy described above, APT data was collected from a specimen which was milled to have a layer of bulk ice near the apex, followed by a layer of nanoporous Cu, as shown in Figure 4(c). Several small fractures occurred over the course of the experiment, so the "bulk ice" region and "nanoporous Cu" parts of the data were reconstructed and analysed separately.

Figure 4 shows the mass spectra from the bulk ice region and the nanoporous Cu region. Prominent peaks have been labelled. In the (a) 0–70 Da range of the nanoporous Cu region, the mass spectrum is dominated by $^{63}Cu^+$ and $^{65}Cu^+$, and protonated water clusters described by $H(H_2O)_n^+$ (or $(H_2O)_{n-1}(H_3O)^+$ [35]) with $n \geq 1$. The tallest of these is the $n = 1$ (i.e. $H(H_2O)^+$), though the $n = 2$ and $n = 3$ are visible. Larger clusters with $n \geq 4$ are small or not observed. $O^+$, $OH^+$ and $H_2O^+$, are respectively observed at 16, 17 and 18 Da. Other notable peaks include $Mn^{++}$ (27 Da), $O_2^+$ (32 Da), and a peak at 36 Da ranged as $HCl^+$ [56]. The observation of prominent $H_2O$ peaks in the mass spectrum confirms that ice was present in the pores of the nanoporous Cu region.

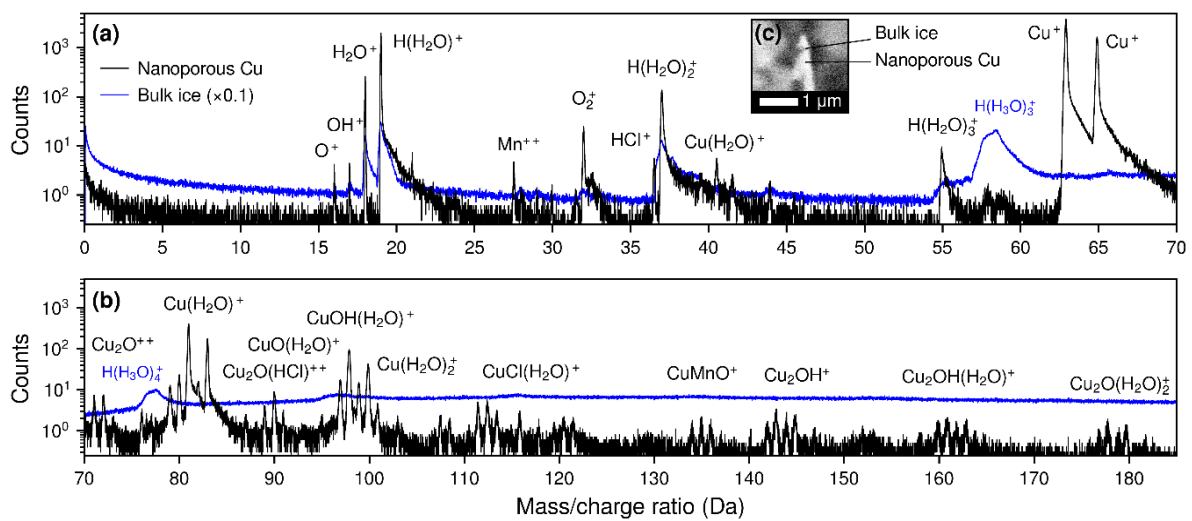

Figure 4: Mass spectra from the nanoporous Cu region (black) and the bulk ice region (blue), from (a) 0 − 70 Da and (b) 70 − 185 Da. Peaks found only in the bulk ice region are labelled in blue text, otherwise peak labels refer to the mass spectrum of the nanoporous Cu region. (c) Combined TLD secondary electron and backscattered electron image of the specimen analysed by APT. Annotations indicate the nanoporous Cu and bulk ice regions.

A more complex mass spectrum is observed in the (b) 70 – 185 Da range. For clarity, only the most prominent peaks are labelled. All peaks in this range are molecular ions with varying stoichiometry of Cu, Mn, O, OH, $H_2O$, $H(H_2O)$, and Cl, with the most abundant species being $Cu(H_2O)^+$ and $CuOH(H_2O)^+$. Although prominent in the mass spectrum, these complex species do not make up a significant fraction of the overall sample composition. Figure 5 shows the bulk composition of the nanoporous Cu region in ionic, atomic and molar fraction. 92.47 ion.% of the sample is comprised of Cu, $H_2O$, $Cu(H_2O)$, and $H(H_2O)_n$ up to $n \leq 3$. If it's assumed that all the O in this region of the sample originated from $H_2O$, then the molar fraction of $H_2O$ is 37.7 mol.%, with an excess H of 8.31 mol.%. This excess H is likely due to adsorption from the residual $H_2$ in the analysis chamber [57]. No peaks at 1 Da or 2 Da are observed despite the relatively slow evaporation rate (125 kHz pulse rate, 1.0% target detection rate).

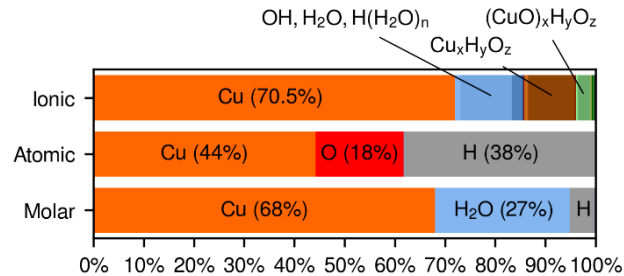

Figure 5: Bulk composition of the nanoporous Cu + $H_2O$ reconstruction, divided into ionic, atomic and molar fractions. Colour-coding indicates Cu (orange), O (red), H (grey), OH, $H_2O$ and $H(H_2O)_n$ (blues), $Cu_xH_yO_z$ species (browns) and $(CuO)_xH_yO_z$ (greens) species. The fraction of other species (such as Mn, Cl) are too small to be seen at this scale, and are assumed to be negligible for the calculation of molar fractions.

The mass spectrum for the bulk ice region is included for comparison with prior reports on the study of ice [33,35]. It is dominated by broad peaks of $H(H_2O)_n^+$ up to $n \leq 3$ and $H(H_3O)_n^+$ with $n \geq 3$, with a particularly strong thermal tail following the $n = 3$ peak at 58 Da. Although it may be possible to improve the quality of the mass spectrum from the bulk ice region by optimising acquisition conditions [33], this was not done here.

Figure 6 shows the reconstructed volume of the nanoporous Cu data, visualised by (a) isosurfaces of Cu at 85 at.% (orange) and (OH, $H_2O$, $H(H_2O)_n$, $Cu(H_2O)$) at 15 at%. (transparent blue), and reconstructed ion positions for (b) $Cu^+$, (c) $H(H_2O)_n^+$ with $1 \leq n \leq 3$, (d) $Cu(H_2O)^+$, and (e) $CuOH(H_2O)^+$. The isosurfaces and reconstructed ion positions depict what is most likely the edge of a Cu ligament surrounded by a $H_2O$-filled pore. Figure 7 shows a one-dimensional concentration profile over the ligament-pore interface in the reconstruction $xy$ plane. The $H_2O$ content in the ligament is small but measurable. The interface region is dominated by $H_2O^+$ and $H(H_2O)^+$, with larger clusters becoming slightly more dominant further from the surface. The presence of $H_2O$ in the ligament and Cu in the pore is likely the result of ion trajectory aberrations arising from the differing evaporation fields of Cu (30 V/nm [47]) and $H_2O$ ($\approx 10$ V/nm [39]).

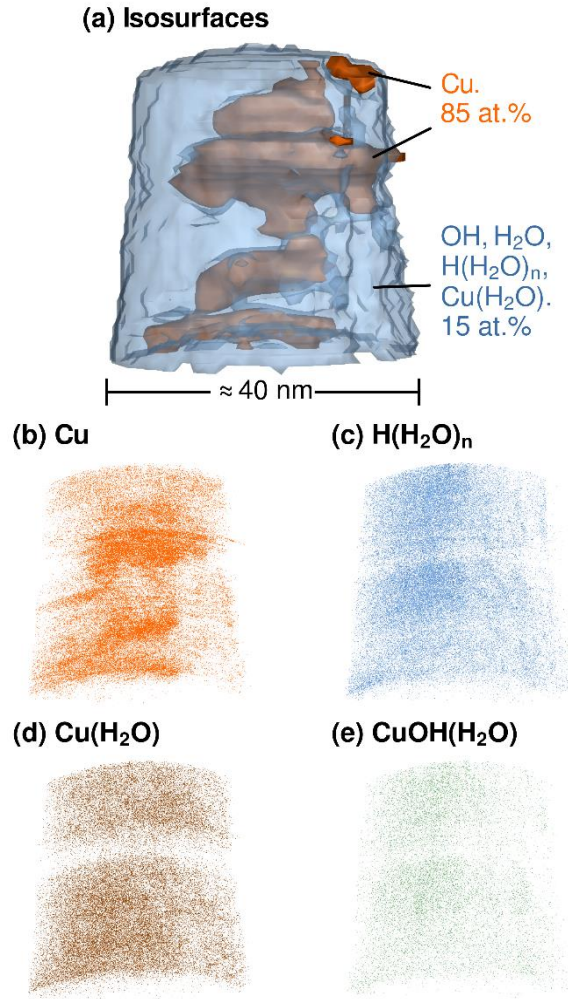

Figure 6: Reconstructed volume from the nanoporous Cu region, showing (a) isosurfaces of Cu at 85 at.% and OH, $H_2O$, $H(H_2O)_n$ ($1 \leq n \leq 3$) and $Cu(H_2O)$ at 15 at.%, and reconstructed ion positions of (b) Cu, (c) $H(H_2O)_n$ ($1 \leq n \leq 3$), (d) $Cu(H_2O)$ and (e) $CuOH(H_2O)$. Reconstruction width and height are both ≈40 nm, but no spatial calibration was performed so distances should be considered approximate.

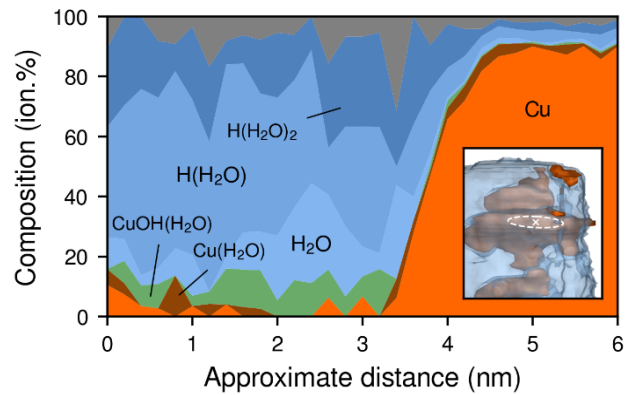

Figure 7: A one-dimensional concentration profile over the ligament-pore interface shown in Figure 6. The profile has been calculated in the reconstruction $xy$ plane, as indicated by the inset image, with the × indicating that the distance along the profile is increases into the page. Individual profiles are stacked and labelled, and the grey is the sum of the profiles of all other species.

Figure 8(b) shows the multiples histogram (or "Saxey plot") [58] for the nanoporous Cu region. The corresponding mass spectrum is shown in (a). The dramatic $H_2O_n$ disassociation tracks reported by Schwartz et al. [35] were not observed in this work. Most of the multiple events observed here are of Cu and $H_2O$ complexes. With reference to the annotations in Figure 8, the dominant disassociation tracks are:

$$Cu(H_2O)^+ \rightarrow Cu^+ + H_2O \qquad (1)$$

$$(Cu_2O)(H_2O)^{++} \rightarrow Cu^+ + (CuO)(H_2O)^+ \qquad (2)$$

$$Cu(H_2O)_2^+ \rightarrow Cu(H_2O)^+ + H_2O \qquad (3)$$

Equations 1 and 3 produce neutral molecular $H_2O$, resulting in dissociation tracks where one product has infinite mass-to-charge ratio. The multiples histogram also shows measurable co-evaporation of $O_2^+$ (32 Da) with $Cu^+$ (63, 65 Da) and $Cu(H_2O)^+$ (81, 83 Da), suggesting significant complex ion evaporation and disassociation from a Cu-O-$H_2O$ chemical environment.

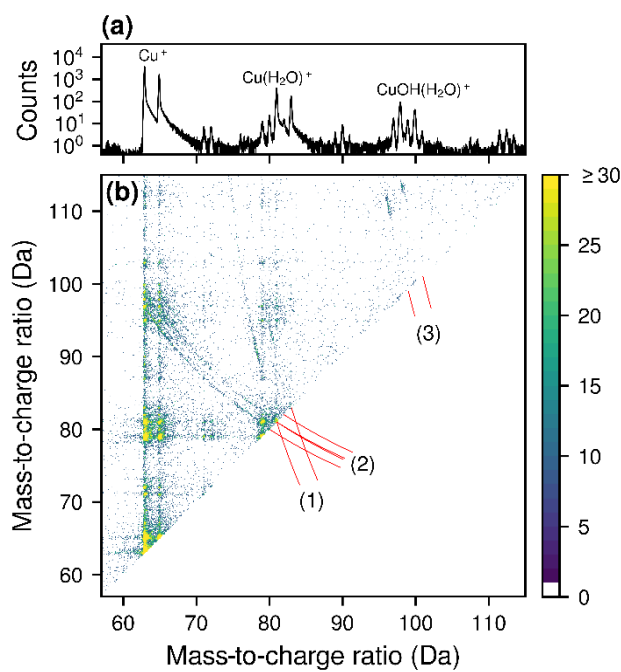

Figure 8: (a) Selected mass spectrum from the nanoporous Cu region, with (b) corresponding multiples histogram (or "Saxey plot") [58]. Prominent peaks are labelled. Selected molecular disassociations are indicated with solid red lines and equation numbers. No disassociation lines are observed to start below 63 Da.

## 4  Discussion

This manuscript describes a workflow for the study of water ice using APT, with the expectation that the method can likely be applied to study any aqueous solution or suspension. One of the major advantages of this workflow to those reported previously is the simplicity of the mass spectrum below 70 Da. Figure 9 compares the mass spectrum from this work to those reported by El-Zoka et al. [33], Exertier et al. [59] and Schwarz et al. [35]. Although Schwarz et al. used a bespoke atom probe [60], El-Zoka et al. used the same atom probe as the present study, and Exertier et al. used a similar LEAP 4000 X Si. El-Zoka et al. and Schwarz et al. studied bulk ice, whereas Exertier et al. studied water adhered to metallic W. The mass spectrum from our work is dominated by $H(H_2O)^+$, $Cu^+$, and $Cu(H_2O)^+$. Larger clusters such as $H(H_2O)_n^+$ for $n \geq 3$ are in very low abundance or absent. Other $H_2O$-$OH$ clusters such as $(H_2O)_n(OH)_m^{+,++}$ [35] are not observed at all. Figure 10 shows the relative abundance of $n$ in $H(H_2O)_n^+$ between $1 \leq n \leq 5$ (where $n = 1$ includes $H_2O^+$) from the four mass spectra shown in Figure 9. Annotations show the average number of water molecules per evaporated ion (i.e., $\bar{n}$) to range from 1.11 in Exertier et al., 1.21 in this work, and up to 2.99 for El-Zoka et al. With fewer peaks in this range of the mass spectrum there is reduced likelihood of peak overlap with solute ions, which is important for future studies into aqueous solutions of organic or biological materials.

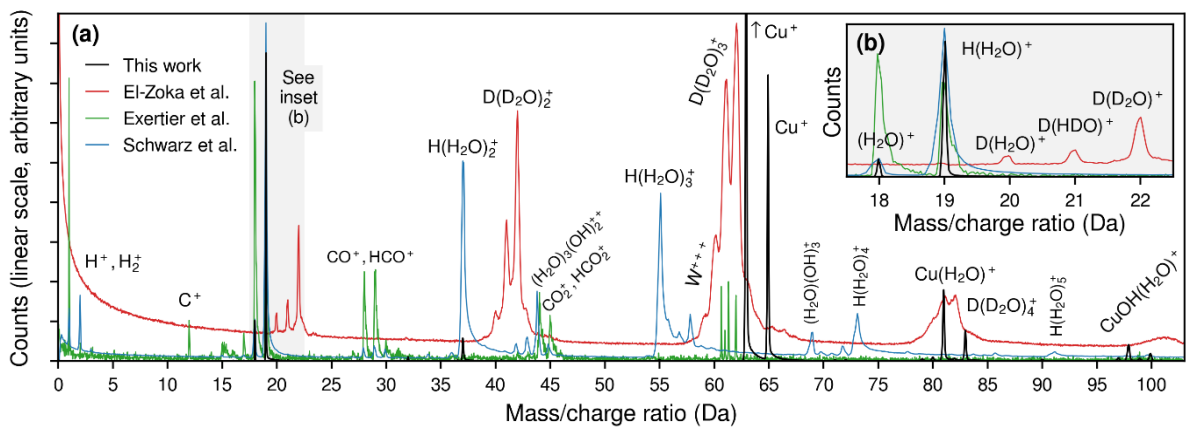

Figure 9: Mass spectra of water ice from this work El-Zoka et al. [33], Exertier et al. [59] and Schwarz et al. [35], with (a) showing 0–105 Da and the inset axes (b) showing 18–22 Da around $H(H_2O)^+$. Vertical axis has a linear scale and mass spectra have been arbitrarily scaled for easily comparison of features. The $^{63}Cu^+$ peak in the mass spectrum from this work extends beyond the upper limit of the vertical axis.

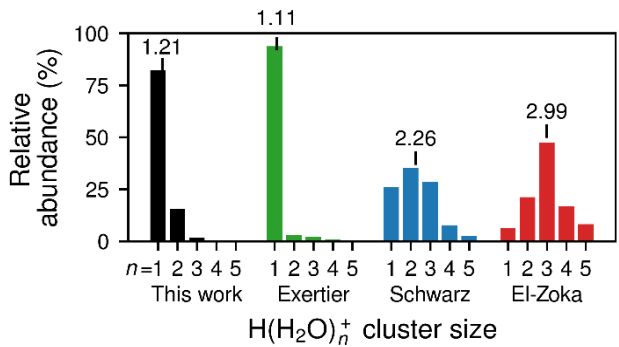

Figure 10: Approximate relative abundance of $H(H_2O)_n^+$ ions between $1 \leq n \leq 5$ (where $n = 1$ also includes $H_2O^+$) for this work, Exertier et al. [59], Schwarz et al. [35] and El-Zoka et al. [33]. For the El-Zoka et al. data, all combinations of H/D (e.g $H(D_2O)^+$, $D(H_2O)^+$) were summed according to the number of $(H/D)_2O$ molecules. Annotations indicate the average cluster size based on the histograms shown.

Another distinctive feature of this work is the greatly reduced peak widths compared to previous studies. This is most evident around the $H(H_2OF)^+$ peak at 19 Da, as seen in the inset axes Figure 9(b). These findings are most likely the result of the nanoporous Cu framework extending all the way to the apex. As nanoporous metals prepared via dealloying should have fully bicontinuous pores and ligaments [48]; it is likely that any Cu ligaments evaporating at the specimen apex have full electrical contact with the sample stage. This will lead to delivery of voltage all the way to the specimen apex, resulting in enhanced electric field, leading to generally smaller $H(H_2O)_n^+$ clusters [61]. This should also result in better dissipation of heat from the laser pulses, leading to reduced thermal tails and better mass resolving power. The full-width half-maximum ($\sigma$) of the $H(H_2O)_n^+$ $n = 1$ and $n = 2$ peaks in Figure 4 are $\sigma_{n=1} = 0.050 \pm 0.001$ Da and $\sigma_{n=2} = 0.122 \pm 0.003$ Da in the nanoporous Cu region, and $\sigma_{n=1} = 0.206 \pm 0.004$ Da and $\sigma_{n=2} = 0.61 \pm 0.02$ Da in the bulk ice region. As peak widths are ≈4–5× greater in the bulk ice region, this indicates higher cooling rates in the nanoporous Cu region [2].

The goal of this project is to use nanoporous needles as a support to study solutions of organic materials, or biomaterials such as proteins, by cryo-APT. However, the use of a nanoporous metal framework will naturally result in the metal appearing in the mass spectrum [62]. Cu was observed predominantly as $Cu^+$ (63, 65 Da) which does not significantly overlap with key peaks from prior studies into organic materials like glucose [36], honey [37], *n*-Tetradecane [38], ferritin [16], or immunoglobulin G [15]. We also observed significant Cu-O-($H_2O$) complex ion peaks, particularly $Cu(H_2O)^+$ (81, 83 Da) and $CuOH(H_2O)^+$ (98, 100 Da). One explanation for the prominence of these ions is oxidation of the Cu surface from the dealloying solution. Another is that adsorbed $H_2O$ on the Cu ligament surface (pristine or oxidised) disassociates to form species such as O or OH [63,64], which may further disassociate after field evaporation. Both scenarios are consistent with the reconstructed volumes and the multiples histogram shown here. Although these Cu-O-($H_2O$) peaks complicate the reconstruction, their high mass-to-charge ratios mean they are also unlikely to overlap with significant molecular fragments from organic materials described above. Optimisation of sample preparation (e.g. dealloying conditions such as acid concentration) or acquisition conditions (e.g. laser pulse energy or pulse rate) could reduce the prominence of these peaks.

The difference in evaporation field between the Cu (30 V/nm [47]) and $H_2O$ ($\approx$ 10 V/nm [39]) will result in local magnification, distorting the reconstructed volume. This is likely the cause of the significant $H_2O$ content in the Cu ligament, and vice versa, observed in the concentration profile shown in Figure 7. The effect of these distortions on structural studies of molecules in solutions is not known. It may be possible to minimise these distortions using more sophisticated reconstruction protocols [65–67]. The diameter of the pore will also affect the maximum size of the molecule which can be analysed, and the metal surface may also lead to adsorption or structural changes in complex molecules such as proteins. The study of organic molecules in solution loaded into a nanoporous metal is an avenue for future work.

## 5   Conclusions

A nanoporous Cu needle was used to support water ice for analysis by APT. Tips were prepared by electropolishing a microstructured $Cu_{0.3}Mn_{0.7}$ alloy, dealloying to produce nanoporous Cu, plunge-freezing in liquid nitrogen, sharpening with cryo-PFIB, and cryotransfer to the atom probe. Regions of bulk ice and nanoporous Cu + $H_2O$ were observed, and the latter was studied in detail. The reconstructed volume appears to show the edge of a Cu ligament extending into a $H_2O$-filled pore. The mass spectrum (below 70 Da) is dominated only by $H(H_2O)^+$ and $Cu^+$, though we observe significant evaporation of Cu-O-$H_2O$ complex ions above 70 Da. We also observe probable local magnification due to large differences in evaporation field between the Cu and $H_2O$.

This work is an advance in the study of frozen water using APT. Techniques for the electropolishing, dealloying, and careful FIB-milling of nanoporous Cu needles have been developed. The continuous Cu framework increases the electrical and thermal conductivity of the sample, increasing local electric field intensity at the apex, minimising the evaporation of large ($n \geq 2$) $H(H_2O)_n^+$ clusters, and minimising sample heating from the pulsed laser. These effects simplify the mass spectrum, minimising the chances of a problematic peak overlap with important organic fragments in the study of organic molecules in solution. Still, many parts of the workflow can be optimised, including in material design, electropolishing and dealloying parameters, data acquisition parameters in the atom probe, and protocol choices in data reconstruction. This sample preparation method could potentially be combined with a graphene [26,29] or metallic coating [68], removing the need for a fully cryogenic workflow.

## Acknowledgements

L. Tegg, and J. M. Cairney acknowledge the Australian Research Council grant FT180100232. The authors acknowledge the technical and scientific assistance of Sydney Microscopy & Microanalysis (SMM), the University of Sydney node of Microscopy Australia, and the facilities and assistance of the Sydney Manufacturing Hub, a Core Research Facility of University of Sydney. The authors particularly thank T. Sato, A. I. K. Pillai and V. Bhatia of SMM for experimental support and insightful discussion. E. V. Woods, S-.H. Kim, A. A. El-Zoka, and B. Gault are grateful for funding from the ERC for the project SHINE (ERC-CoG) #771602. R. Dubosq, E. V. Woods and B. Gault are grateful to the DFG for funding through the Leibniz Award 2020. S-.H. Kim, A. A. El-Zoka and B. Gault are grateful to the DFG through the DIP Project No. 450800666. We thank U. Tezins, C. Broß and A. Sturm for their support at the FIB and APT facilities at Max-Planck-Institut für Eisenforschung (MPIE). T. M. Schwarz of MPIE is thanked for providing data for inclusion in Figure 9.